\documentclass[prd,amsmath,amssymb,a4paper,twocolumn]{revtex4}
\usepackage{hyperref,epsfig,graphics}

\begin{document}

\title{Chemical freeze-out in heavy ion collisions at large baryon densities}
\author{Stefan Floerchinger$^1$ and Christof Wetterich$^2$}
\affiliation{$^1$Physics Department, Theory Unit, CERN, CH-1211 Gen\`eve 23, Switzerland\\
$^2$Institut f\"{u}r Theoretische Physik, Universit\"at Heidelberg, Philosophenweg 16, D-69120 Heidelberg}

\begin{abstract}
We argue that the chemical freeze-out in heavy ion collisions at high baryon density is not associated to a phase transition or rapid crossover. We employ the linear nucleon-meson model with parameters fixed by the zero-temperature properties of nuclear matter close to the liquid-gas quantum phase transition. For the parameter region of interest this yields a reliable picture of the thermodynamic and chiral properties at non-zero temperature. The chemical freeze-out observed in low-energy experiments occurs when baryon densities fall below a critical value of about 15 percent of nuclear density. This region in the phase diagram is far away from any phase transition or rapid crossover. 
\end{abstract}

\maketitle

Relativistic heavy ion collisions are a promising way to investigate the properties of fundamental quantum field theory -- more specific QCD -- at nonzero temperature and density. However, in contrast to most condensed matter systems the produced matter has only a short live time during which it expands and cools quickly. It is a big challenge to determine the properties of the produced matter. Basically all information has to be reconstructed from the final state, i.\ e.\ from the momenta and chemical composition of the detected particles.

Important observables are the yields of different particle species. To surprisingly good approximation they can be described by the so called statistical model which assumes a thermal distribution of a non-interacting hadron resonance gas \cite{StatisticalModel, ABMS2006, ABMS2009}. The most prominent observables extracted from the thermal fits are the temperature and baryon chemical potential associated with the chemical freeze-out. 

It has been advocated that the chemical freeze-out temperature coincides with the temperature of the QCD phase transition for low baryon density \cite{BMSW}. The basic argument states that the rates of particle number changes in the hadronic phase are too small to maintain chemical equilibrium. This holds whenever scattering processes involving only a few particles dominate. Chemical freeze-out therefore occurs for particle number densities that are just high enough that multi-particle scattering or collective effects dominate. For low baryon number that is possible only for temperatures very close to a phase transition or crossover. The difference between the critical temperature and the freeze-out temperature has been estimated to be less than $5\; \text{MeV}$. The measured freeze-out temperature at RHIC at heavy ion collisions with $\sqrt{s_{NN}}=200\;\text{GeV}$ of $164\pm 6\;\text{MeV}$ \cite{ABMS2009} agrees actually rather well with the critical temperature of the crossover \cite{latticeCrossover} in lattice simulations of $T_c \approx (157\pm 10)\;\text{MeV}$ \cite{lattice1} and $T_c \approx (154\pm 9)\;\text{MeV}$ \cite{lattice2}.

The agreement between the critical temperature of a phase transition or crossover and the observed chemical freeze-out temperature is expected to hold for low values of the chemical potential $\mu$ in the QCD phase diagram. Naturally, the question arises whether a similar argument can be extended to higher values of $\mu$ or larger baryon density. Interesting ideas for this issue have been proposed recently \cite{BL}, where the chemical freeze-out was connected with the hypothetical transition to quarkyonic matter. We ask: does the curve of measured freeze-out temperatures reflect a phase transition line or rapid crossover in the whole $\mu$-$T$-plane?

\begin{figure}
\centering
\includegraphics[width=0.5\textwidth]{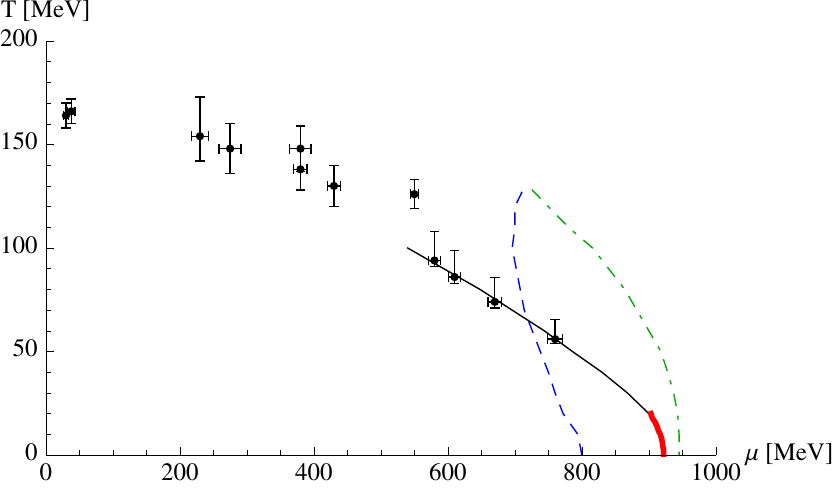}
\caption{Curve of constant baryon number $n_\text{Baryons}=0.15\;n_\text{nuclear}$ in the Meson-Baryon model (solid black line). The points with error-bars mark the chemical freeze-out as obtained from the fits to experimentally measured particle yields \cite{ABMS2009}. The red line marks the first order phase transition to nuclear matter. The dashed and dashed-dotted lines indicate an estimate for the range of applicability of our model. More specific, in the region to the right of the dashed line the relative contribution of pions to the pressure is smaller than $20\%$. In the region to the left of the dashed-dotted line the baryon density $n_\text{Baryons}$ is smaller than 1.5 times the nuclear saturation density $n_\text{nuclear}=0.153/\text{fm}^3$. In this region no signs of a phase transition are visible.}
\label{fig:1}
\end{figure}

\begin{figure}
\centering
\includegraphics[width=0.5\textwidth]{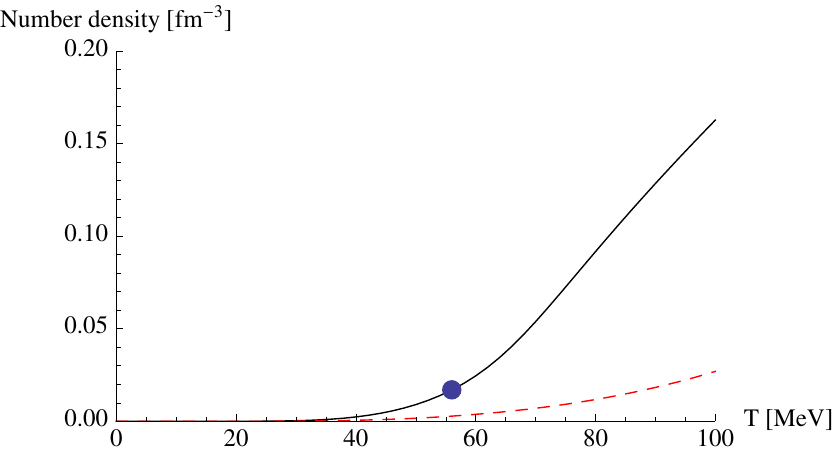}
\caption{Number density of baryons as a function of the temperature for $\mu=750\;\text{MeV}$ (solid line). Note that the number of anti-baryons is negligible within the plot resolution. We also show the number of pions (dashed line). The dot marks the experimental result for the chemical freeze-out temperature $T_\text{ch}=56^{+9.6}_{-2.0}\;\text{MeV}$ corresponding to $\mu_\text{ch}= 760\pm22.8\,\text{MeV}$.}
\label{fig:2}
\end{figure}

\begin{figure}
\centering
\includegraphics[width=0.5\textwidth]{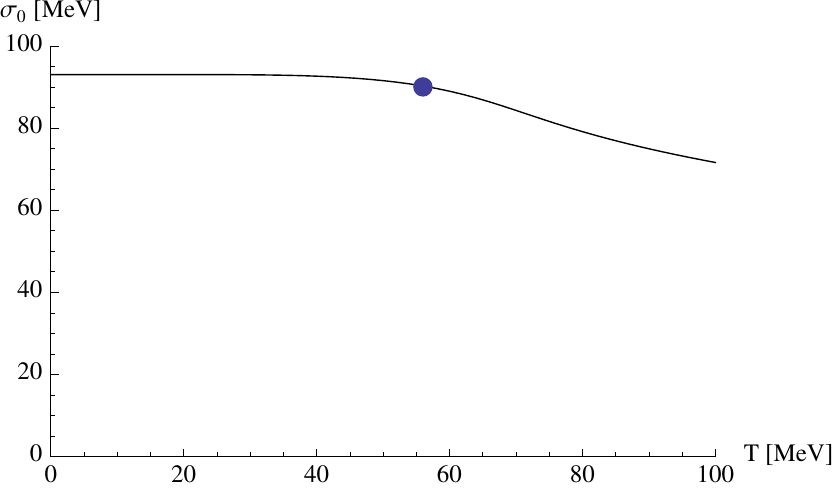}
\caption{Chiral order parameter as a function of the temperature for $\mu=750\;\text{MeV}$. The dot marks the experimental result for the chemical freeze-out temperature $T_\text{ch}=56^{+9.6}_{-2.0}\;\text{MeV}$ corresponding to $\mu_\text{ch}= 760\pm22.8\,\text{MeV}$.}
\label{fig:3}
\end{figure}

In this letter we argue that this is not the case. The observed freeze-out temperatures $T_\text{ch}$ for the largest values of $\mu$ lie actually in a region where a simple modeling by baryons and mesons becomes possible. While a rapid change of the particle density with temperature continues to play a crucial role for the determination of $T_\text{ch}$, this actually happens in a region that is substantially away from any transition or crossover. We illustrate the situation in Fig.\ \ref{fig:1} where we indicate the observed points in the $\mu$-$T$-plane and the region of validity of a simple baryon-meson model. We also demonstrate in Fig.\ \ref{fig:2} the change of particle density with temperature for fixed value of $\mu$. The dot in this figure indicates the measured value of $T_\text{ch}$ for $\mu=760\;\text{MeV}$. A similar figure \ref{fig:3} for the chiral order parameter $\sigma_0$ as a function of $T$ shows that no particular distinct feature such as a (chiral) phase transition or crossover is visible in this range. The deviations of $\sigma_0$ from the vacuum value are small in the whole range of the black solid curve in Fig.\ \ref{fig:1}.

For a discussion of the phase diagram and the thermodynamic and chiral properties within the region indicated in Fig. \ref{fig:1} the linear nucleon-meson model is a reasonable approximation. It uses as degrees of freedom the proton and neutron, a neutral vector meson $\omega_\mu$ as well as the pions and the collective $\sigma$-meson. (We ignore isospin violation and electromagnetism for simplicity.) Chiral symmetry is implemented explicitly. Integrating out the $\sigma$-degree of freedom yields an effective non-linear $\sigma$-model coupled to nucleons and the vector meson. For this latter model the chiral perturbation theory has been used extensively \cite{ChiralPT,CPT2,CPT3}. On the other hand, within a quadratic approximation to the effective potential of the field $\sigma$, and if only the dominant nucleon fluctuations are included for the computation of the chiral order parameter and the baryon density, one recovers the gap equations of the Walecka model \cite{WaleckaModel}. Parameters of the effective potential at zero temperature and the chemical potential $\mu_c$ corresponding to the gas-liquid phase transition in nuclear matter can be determined from observation. In this parameter region the model can be mapped directly to the nuclear droplet model.

The computational task concerns then mainly the difference of the effective meson potential $U(\sigma; T,\mu)-U(\sigma; 0, \mu_c)$. This can be done by various methods -- for example one could employ functional renormalization by adding nucleon degrees of freedom to the setting of ref.\ \cite{Jungnickel:1996fd}. For our limited purpose a very simple approach will do. The potential difference is directly related to difference of pressure for the parameters $(\sigma; T,\mu)$ and $(\sigma; 0,\mu_c)$. This can be approximated by a free gas of nucleons with $\sigma$-dependent mass. We can consider $\sigma$ as an additional parameter in thermodynamics. Its value can be varied by varying the quark mass. If needed, meson fluctuations can be added in a similar way. We will discuss the linear nucleon-meson model in the setting of ref.\ \cite{Berges:1998ha}. (Our normalization of $\sigma$ differs by a factor $2$ from \cite{Berges:1998ha}.) Our new results extend the analysis to non-vanishing temperature. 

We consider our calculation as a reliable estimate of the temperature and density dependence of thermodynamic equilibrium quantities within the region of validity indicated in fig.\ \ref{fig:1}. Input from observations is only used for $T=0$.  An important ingredient is the dependence of particle masses on the chiral order parameter $\sigma$. In turn, $\sigma$ depends on $\mu$ and $T$. This effect goes beyond a resonance gas model with fixed vacuum masses and permits to cover nuclear matter as well. Nearby the gas-liquid phase transition line and its possible continuation beyond the critical endpoint by a crossover line we find that the $\sigma$-dependence of particle masses is quantitatively important and crucial for an understanding of the phase diagram and thermodynamic quantities. It is an important finding of our paper , however, that this effect becomes small for the parameter region of the experimentally realized chemical freeze-out.

\section*{Linear nucleon-meson model}
We use an effective model for baryons $\psi_a$ ($a$ is an isospin index with $\psi_1$ describing protons and $\psi_2$ neutrons), an isospin singlet vector meson $\omega_\mu$, a scalar meson $\sigma$ and pseudo-scalar mesons $\pi^0=\pi_3$, $\pi^\pm=\frac{1}{\sqrt{2}}(\pi_1\pm i \pi_2)$. It is convenient to combine the scalars and pseudo-scalars in the field
\begin{equation}
\phi_{ab} = \begin{pmatrix} \tfrac{1}{\sqrt{2}}(\sigma+i \pi^0) && i \pi^- \\ i \pi^+ && \tfrac{1}{\sqrt{2}}(\sigma-i \pi^0)\end{pmatrix}.
\end{equation}
The effective Lagrangian is of the form
\begin{equation}
\begin{split}
&{\cal L} = \bar \psi_a \; i \gamma^\nu (\partial_\nu -i\, g \;\omega_\nu-i\, \mu \,\delta_{0\nu} ) \;\psi_a\\
&+ \sqrt{2}\, h {\big [} \bar \psi_a \left(\tfrac{1+\gamma_5}{2}\right) \phi_{ab} \psi_b +\bar \psi_a \left(\tfrac{1-\gamma_5}{2}\right) (\phi^\dagger)_{ab} \psi_b {\big ]} \\
& + \tfrac{1}{2} \phi^*_{ab} (-\partial_\mu \partial^\mu) \phi_{ab} + U_\text{mic}(\rho,\sigma) \\
& +\frac{1}{4} (\partial_\mu \omega_\nu - \partial_\nu \omega_\mu) (\partial^\mu \omega^\nu-\partial^\nu \omega^\mu) + \frac{1}{2} m_\omega^2\, \omega_\mu \omega^\mu.
\end{split}
\label{eq:effmodel}
\end{equation}
Here we use the chiral invariant scalar field combination $\rho=\tfrac{1}{2}\phi^*_{ab} \phi_{ab}$ and $U_\text{mic}(\rho,\sigma)$ is a microscopic form of the effective potential
\begin{equation}\label{2A}
U_\text{mic}(\rho,\sigma)=\bar U(\rho)-m^2_\pi f_\pi\sigma.
\end{equation}
The Lagrangian \eqref{eq:effmodel} is invariant under the chiral symmetry $SU(2)_V \times SU(2)_A \times U(1)_V \times U(1)_A$ where the nucleon doublet transforms according to
\begin{equation}
\begin{split}
\psi & \to \left( 1+ \frac{i}{2} \boldsymbol{\alpha}_V \boldsymbol{\tau} + \frac{i}{2} \boldsymbol{\alpha}_A \boldsymbol{\tau} \gamma_5 + \frac{i}{2} \beta_V + \frac{i}{2} \beta_A \gamma_5 \right) \psi,\\
\bar \psi & \to \bar \psi \left( 1- \frac{i}{2} \boldsymbol{\alpha}_V \boldsymbol{\tau} + \frac{i}{2} \boldsymbol{\alpha}_A \boldsymbol{\tau} \gamma_5 - \frac{i}{2} \beta_V + \frac{i}{2} \beta_A \gamma_5 \right),
\end{split}
\end{equation}
where $\boldsymbol{\tau}=(\tau_1,\tau_2,\tau_3)$ are the Pauli matrices in isospin space and the scalar field transforms according to
\begin{equation}
\phi \to \phi - \frac{i}{2} \boldsymbol{\alpha}_V [\boldsymbol{\tau}, \phi] - \frac{i}{2} \boldsymbol{\alpha}_A \{\boldsymbol{\tau},\phi\} + i \beta_A \phi.
\end{equation}
The vector meson field $\omega_\mu$ is invariant under the above symmetry. It couples to the conserved baryon number current associated with the $U(1)_V$ symmetry such that only the spin-one part of $\omega_\mu$ plays a role while the spin-zero component $\partial_\mu \omega^\mu$ decouples.

The only explicit breaking of chiral symmetry comes from the quark masses. This is reflected by the linear term in the effective potential \eqref{2A}. Therefore $U_\text{mic}$ depends explicitly on the field $\sigma$ in addition to the invariant $\rho=\tfrac{1}{2}(\sigma^2+\boldsymbol{\pi}^2)$. The scalar field $\sigma$ will have a vacuum expectation value, in contrast to the pseudo-scalar field, $\pi^0=\pi^+=\pi^- = 0$. Due to rotational symmetry, only the zero-component of the vector field $\omega_\mu$ can have an expectation value. Inspection of Eq. \eqref{eq:effmodel} shows that this can be interpreted as a shift in the effective chemical potential.

We are interested in the quantum effective potential $U(\sigma,\omega_0)$ which includes the effects of quantum and thermal fluctuations. It can be obtained from the quantum effective action - the generating functional for one-particle irreducible Greens functions - by specializing to constant $\sigma$ and $\omega$ with $\psi=0,\pi=0$. The minimum of $U(\sigma,\omega_0)$ determines the expectation values for $\sigma$ and $\omega_0$, i.e. the chiral order parameter and the effective chemical potential. In general, the computation of $U(\sigma,\omega_0)$ from a microscopic action is a complicated task. From the realization of symmetries we know, however, that the explicit symmetry breaking occurs only via the unaffected linear term,
\begin{equation}\label{F2a}
U(\sigma,\omega_0)=U(\rho,\omega_0)-m^2_\pi f_\pi\sigma~,~\rho=\frac12\sigma^2,
\end{equation}
such that the task consists in a computation of the chirally invariant potential $U(\rho,\omega_0)$. 

We are only interested in the difference $\Delta=U(\rho,\omega_0;T,\mu)-U(\rho,\omega_0;0,\mu_c)$, with $\mu_c$ the value of the chemical potential at which the zero-temperature phase transition between a hadron gas and nuclear matter occurs. This simplifies our task considerably. Instead of a complicated computation of $U(\rho,\omega_0;0,\mu_c)$ we can use observation in order to pin down the relevant properties of this quantity. In a functional renormalization approach the difference $\Delta$ involves only mesons with mass $m$ smaller $\pi T$ or baryons with $m-\mu_\text{eff}$ smaller than $\pi T$, where $\mu_{eff}=\mu+g\bar\omega_0$ and $\bar\omega_0$ is the expectation value of $\omega_0$  \cite{Jungnickel:1996fd,Berges:1998ha}. In our range of interest these are essentially nucleons and possibly pions, thus justifying the degrees of freedom incorporated in our model.

Furthermore, for the relatively narrow range in $T$ and $\mu$ that we investigate here the running and the associated $\mu$- and $T$-dependence of the couplings $h$ and $g$ is small and can be neglected. In a first approach we also neglect the subleading pion fluctuations. With these approximations the solution of functional flow equations actually reduces to performing a Gaussian functional integral over the fermionic fields $\psi_N$ in a background of constant $\sigma$ and $\omega_0$. This is nothing else than relativistic mean field theory. We stress that mean field theory is generically not expected to give reliable results in our setting with strong interactions. For example, a mean field computation of $U(\rho,\omega_0;0,\mu_c)$ would fail badly. However, the more general view from a functional renormalization perspective permits to asses that the mean field result for $\Delta$ is reliable within an appropriate parameter range. We expect leading corrections from the omitted pion fluctuations. (They can be incorporated, in principle, in some type of extended mean field theory, see below.) To the right of the left dashed line in Fig. \ref{fig:1} the pion contributions to the pressure are less than 20 \%. A second type of correction is expected due to the neglected $\sigma$-dependence of $h,g$ and $m_\omega$. In view of the small deviation of $\sigma$ from its vacuum value visible in Fig. \ref{fig:3} we expect this effect to be small. It increases, however, for larger density and this is one of the restrictions for the limitation of validity of our model indicated by the right dashed line in Fig. \ref{fig:1}.

We next discuss the mean field contribution to $\Delta$ which is directly related to the pressure of a free nucleon gas with field dependent masses. The corresponding contribution to the effective potential depends on the temperature $T$ and chemical potential $\mu$. It can be parametrized in terms of the pressure of a free gas of relativistic fermions and corresponding antiparticles
\begin{equation}
\begin{split}
& P_\text{FG}(T,\mu,m) = \frac{1}{3} \int \frac{d^3p}{(2\pi)^3}  \frac{\vec p^2}{\sqrt{\vec p^2+m^2}} \\
& \times\left[ \frac{1}{e^{\tfrac{1}{T}(\sqrt{\vec p^2+m^2}-\mu)}+1} + \frac{1}{e^{\tfrac{1}{T}(\sqrt{\vec p^2+m^2}+\mu)}+1} \right].
\end{split}
\label{eq:PFG}
\end{equation}
Within our model, the effective potential for the bosonic fields reads now
\begin{equation}
\begin{split}
U(\sigma,\omega_0;T,\mu) = & U_\text{vac}(\sigma,\omega_0)\\
&- 4\; P_\text{FG}(T,\mu+g\omega_0, h\sigma),
\end{split}
\label{eq:effpotentialcomb}
\end{equation}
where the factor $4$ accounts for the degeneracy in spin and isospin.
For the effective potential in the vacuum (i.\ e.\ at $T=\mu=0$) we use the parametrization
\begin{equation}\label{CF9}
\begin{split}
U_\text{vac}(\sigma, \omega_0) = &\frac{1}{2}m_\pi^2 (2\rho-f_\pi^2) + \frac{1}{8} \lambda (2\rho-f_\pi^2)^2 \\
&+\frac{1}{3} \frac{\gamma_3}{f_\pi^2} (2\rho-f_\pi^2)^3 +\frac{1}{4} \frac{\gamma_4}{f_\pi^4} (2\rho-f_\pi^2)^4 \\
&- m^2_\pi f_\pi (\sigma-f_\pi) - \frac{1}{2} m_\omega^2 \omega_0^2.
\end{split}
\end{equation}
For $T=0$ the pressure $P_{FG}(0,\mu,m)$ vanishes identically for $\mu<m$. Thus $\Delta$ vanishes for $\mu<h\sigma-g\omega_0$ and the expressions in Eq.\ \eqref{eq:effpotentialcomb} and \eqref{CF9} coincide.

\section*{Parameters}
At this point the open parameters of the model are
\begin{equation}\label{CF11}
f_\pi,\; m_\pi,\; \lambda,\; \gamma_3,\; \gamma_4,\; m_\omega, \; h \;\; \text{and}\;\; g.
\end{equation}
We choose the experimentally established values $f_\pi=93\;\text{MeV}$ for the pion decay constant, $m_\pi=135\;\text{MeV}$ for the mass of pions and $m_\omega = 783\;\text{MeV}$ for the mass of the vector meson. The Yukawa coupling $h$ is fixed by the requirement that $h f_\pi$ equals the nucleon mass $m_n=939 \; \text{MeV}$ which gives $h=10$. We have verified that somewhat smaller values of $h$ (cf. ref. \cite{Berges:1998ha}) do not change our results qualitatively.  

The remaining open parameters $g$, $\lambda$, $\gamma_3$ and $\gamma_4$ are fixed by requiring that the model describes at vanishing temperature $T=0$ normal nuclear matter at the gas-liquid phase transition. More specific, the minimum of the effective potential \eqref{eq:effpotentialcomb} is at $\sigma=f_\pi$ for vanishing chemical potential $\mu$ and the baryon density $n_B=-\frac{\partial}{\partial \mu} U$ vanishes at this point. However, for increasing $\mu$ a second minimum at a smaller value of $\sigma$ will develop and a first order phase transition takes place when the two minima are degenerate. From the nuclear binding energy $\epsilon_\text{bind} = - 16.3 \; \text{MeV}$ one can determine the critical chemical potential, $\mu_c = 939 \;\text{MeV}-16.3 \; \text{MeV} = 922.7\; \text{MeV}$.

From minimizing the effective potential \eqref{eq:effpotentialcomb} with respect to $\omega_0$ one finds the self-consistency equation
\begin{equation}
\omega_0 =  - \frac{g}{m_\omega^2}  n_B(0,\mu+g \omega_0, h \sigma)
\label{eq:omega0selfcons}
\end{equation}
with the baryon density
\begin{equation}
n_B(0,\mu+g \omega_0, h \sigma) = 4 \frac{\partial}{\partial \mu} P_\text{FG}(0,\mu+g \omega_0, h \sigma).
\label{eq:baryondens}
\end{equation}
At $\mu_c$ the baryon density jumps fro zero to nuclear saturation density $n_\text{nucl}=0.153/\text{fm}^3$ this shows that also $\omega_0$ changes discontinuously from $\omega_0=0$ to $\omega_{0,\text{nucl}}$ at the first order phase transition.
From the Landau mass $m_L = \sqrt{h^2 \sigma_\text{nucl}^2+p_F^2} = \mu_c + g \omega_{0,\text{nucl}}$ one can determine $g$ and $\omega_{0,\text{nucl}}$. We use $m_L=0.80 \; m_n$ for our numerical calculation, which gives $g=9.5$ and $\omega_{0,\text{nucl}}=-18 \; \text{MeV}$. Using the relation $n_\text{nucl}=\tfrac{4}{6\pi^2} p_F^3$, one can also determine the position of the second minimum of the effective potential with respect to $\sigma$ at the phase transition, $\sigma_\text{nucl} = 69.8 \; \text{MeV}$.

Of the remaining three parameters $\lambda$, $\gamma_3$ and $\gamma_4$ two get fixed by the constraints for a first order phase transition 
\begin{equation}
U(\sigma_\text{nucl},\omega_{0,\text{nucl}};0,\mu_c) = U(f_\pi,0;0,\mu_c)
\end{equation}
and
\begin{equation}
\frac{\partial}{\partial \sigma} U(\sigma_\text{nucl},\omega_{0,\text{nucl}};0,\mu_c) =0.
\end{equation}
The third parameter (say $\lambda$) can now be adapted to other properties of nuclear matter. For example, the choice $\lambda=50$, $\gamma_3=3$ and $\gamma_4=50$ corresponds to the (vacuum) mass of the $\sigma$-meson $m_\sigma = \sqrt{\lambda f_\pi^2 + m_\pi^2} = 670\; \text{MeV}$, the compressibility module $K=9 n / (dn/d\mu) = 300 \; \text{MeV}$ and the surface tension of a nuclear droplet $\Sigma= \int_{\sigma_\text{nucl}}^{f_\pi} \sqrt{2 U(\sigma)} d\sigma = 42 000 \; \text{MeV}^3$.
Considered the simplicity of the model, these values are in reasonable agreement with the experimentally established values $m_\sigma \approx 484 \pm 17\; \text{MeV}$ \cite{sigmaMass}, $K\approx 240\pm30\; \text{MeV}$ and $\Sigma \approx 42200\, \text{MeV}^3$.

As an independent check we determine for this choice of parameters also the ``nuclear $\sigma$-term'' which quantifies how the vacuum mass of the nucleon depends on the chiral symmetry breaking explicit mass term for the quarks. It can be determined by comparing the expectation value $\sigma_\text{nuc}$ in the nuclear matter phase with the value obtained in the chiral limit, $m_\pi=0$, for otherwise identical $U_\text{vac}$. One finds $40\; \text{MeV}$, in reasonable agreement with lattice calculations \cite{sigmaLattice}. Overall, one finds a satisfactory agreement with nuclear matter at vanishing temperature and the nuclear droplet model. Our parameters are found in a similar range as the ones determined in ref. \cite{Berges:1998ha}. We have checked that other reasonable parameter choices do not modify our main conclusions. 

\section*{Thermodynamic properties and chemical freeze-out}
\begin{figure}
\centering
\includegraphics[width=0.5\textwidth]{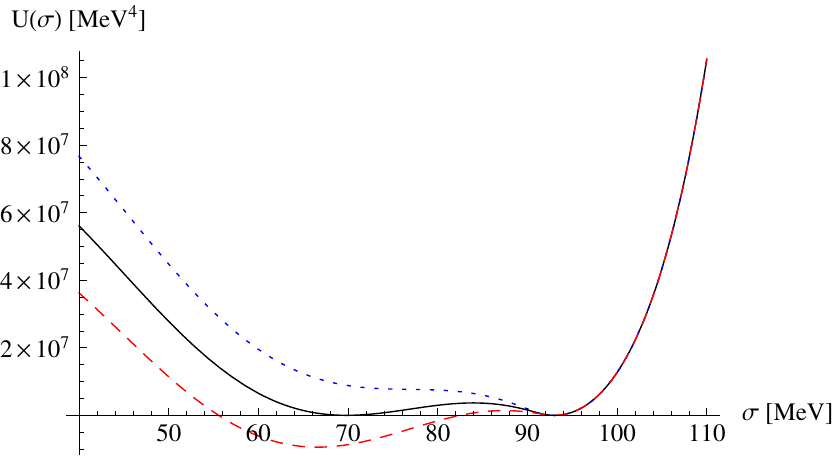}
\caption{Effective potential $U(\sigma)$ as a function of the chiral order parameter for $T=0$ and chemical potential $\mu=915 \;\text{MeV}$ (dotted line), $\mu=922.7\;\text{MeV}$ (solid line) and $\mu=930\;\text{MeV}$ (dashed line).}
\label{fig:EffectivePotentialT0}
\end{figure}

Let us now investigate the linear nucleon-meson model at non-zero temperature. We start from the effective potential $U(\sigma;T,\mu)$ which follows from Eq. \eqref{CF11} by minimizing with respect to the value of $\omega_0$. The complete information about the phase diagram and the thermodynamic properties is encoded  in this quantity. For example, the chiral condensate $\sigma_0(T,\mu)$ is determined as the global minimum of $U(\sigma;T,\mu)$ with respect to $\sigma$. A phase transition of first order occurs when the potential has two local minima, such that at the critical temperature $T_c$ the global minimum jumps discontinuously from one local minimum to the other. To illustrate this behavior we plot in Fig.\ \ref{fig:EffectivePotentialT0} the effective potential $U(\sigma;T,\mu)$ as a function of $\sigma$ for vanishing temperature $T=0$ and different values of the chemical potential $\mu$.
One can clearly see the first order phase transition at the critical chemical potential $\mu_c=922.7\;\text{MeV}$.

\begin{figure}
\centering
\includegraphics[width=0.5\textwidth]{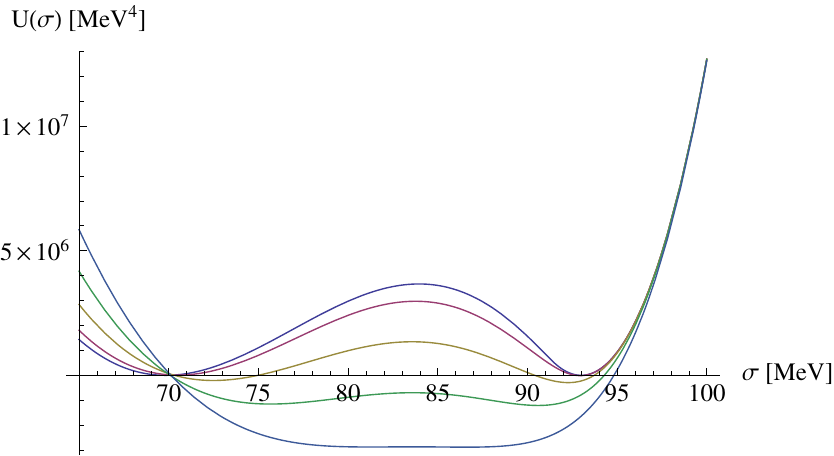}
\caption{Effective potential $U(\sigma)$ as a function of the chiral order parameter at the critical chemical potential of the first order phase transition $\mu=\mu_c(T)$ for temperatures $T=0$, $T=5\,\text{MeV}$, $T=10\,\text{MeV}$, $T=15\,\text{MeV}$ and $T=20\;\text{MeV}$.}
\label{fig:EffectivePotentialmuc}
\end{figure}

Let us now follow the changes in the effective potential as the temperature is increased. In Fig.\ \ref{fig:EffectivePotentialmuc} we plot $U(\sigma)$ for $\mu=\mu_c(T)$ and for different temperatures $T=0$ ($\mu_c=922.7\;\text{MeV}$), $T=5\;\text{MeV}$ ($\mu_c=921.3\;\text{MeV}$), $T=10\;\text{MeV}$ ($\mu_c=917.2\;\text{MeV}$), $T=15\;\text{MeV}$ ($\mu_c=910.8\;\text{MeV}$) and $T=20\;\text{MeV}$ ($\mu_c=902.3\;\text{MeV}$). 
One finds that the effective potential at the minima gets more negative as the temperature increases. This corresponds to an increase in pressure $p=-U_\text{min}$. One also finds that the potential barrier between the two minima becomes smaller, such that the surface tension for a droplet decreases. For $T_*=20.7\; \text{MeV}$ and $\mu=901\;\text{MeV}$ the barrier disappears and the line of first order phase transitions ends in a critical end point. The two minima merge into one. This is completely analoguous to the critical endpoint for the water-vapor transition. The computed temperature for the endpoint, $T_*\approx 20\;\text{MeV}$ agrees well with observation, demonstrating the validity of our treatment of the linear nucleon-meson model.

Besides the first order gas-liquid phase transition, the effective potential \eqref{eq:effpotentialcomb} also exhibits another first order transition at larger values of the chemical potential. Here the chiral condensate jumps to much smaller values which vanish in the chiral limit $m_q=0$. This transition could be associated with a transition from nuclear matter to quark matter and restoration of chiral symmetry. It is obvious that for quark matter the linear nucleon-meson model cannot give a valid description and we therefore cannot trust Eq. \eqref{eq:effpotentialcomb} in this region of the phase diagram. It is possible to modify the model in order to incorporate an effective change from nucleons to quarks. However, the existence and properties of a phase transition depend strongly on details of the change of effective degrees of freedom \cite{Berges:1998ha} such that no reliable information can be gained without a better understanding how nucleons are replaced by quarks. In practice, the chiral condensate in the vicinity of the gas-liquid transition is typically in the range $\sigma=65...93\;\text{MeV}$, while the additional minimum at larger chemical potential occurs at much smaller values, $\sigma < 10\;\text{MeV}$. Obviously, the Taylor expansion of $U_\text{vac}$ for $\sigma$ around $f_\pi$ in Eq. \eqref{CF9} is no longer reliable for such small values of $\sigma$. The investigations in ref. \cite{Berges:1998ha} show that the phase transition to quark matter (if it exists) occurs for baryon densities higher than the ones that would result from the linear nucleon-meson model. Further support for this assessment comes from the analysis of experimental data on radial and elliptic flow \cite{Danielewicz} as well as from calculations employing in-medium chiral perturbation theory \cite{Kaiser}.
Our limitation to baryon densities smaller than $1.5$ times nuclear density, as indicated in Fig. \ref{fig:1}, is therefore a conservative estimate of the validity of our model. 

\begin{figure}
\centering
\includegraphics[width=0.5\textwidth]{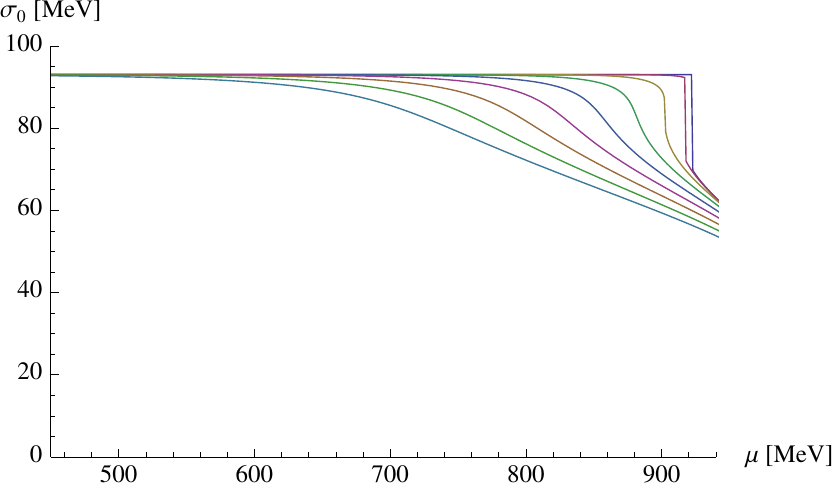}
\caption{Chiral order parameter $\sigma_0$ as a function of the chemical potential for $T=0$ (uppermost curve), $T=10\;\text{MeV}$, $T=20\;\text{MeV}$, $T=30\;\text{MeV}$, $T=40\;\text{MeV}$, $T=50\; \text{MeV}$, $T=60\;\text{MeV}$, $T=70\;\text{MeV}$ and $T=80\;\text{MeV}$ (lowermost curve).}
\label{fig:sigmamin}
\end{figure}
\begin{figure}
\centering
\includegraphics[width=0.5\textwidth]{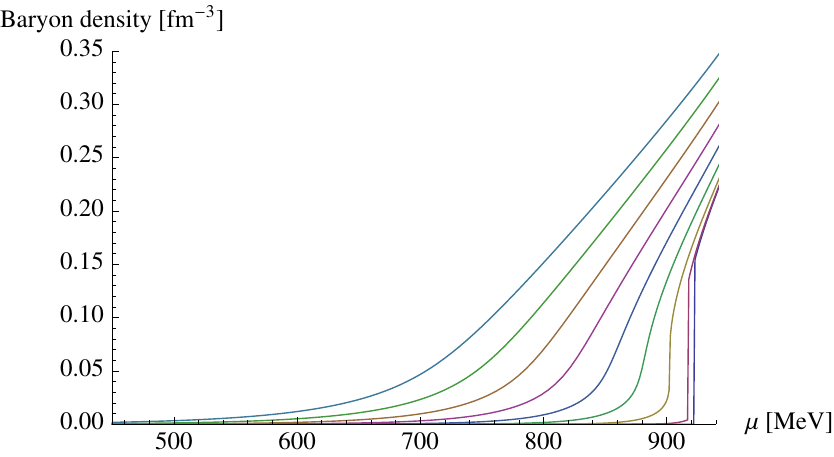}
\caption{Baryon number density as a function of the chemical potential for $T=0$ (lowermost curve), $T=10\;\text{MeV}$, $T=20\;\text{MeV}$, $T=30\;\text{MeV}$, $T=40\;\text{MeV}$, $T=50\; \text{MeV}$, $T=60\;\text{MeV}$, $T=70\;\text{MeV}$ and $T=80\;\text{MeV}$ (uppermost curve).}
\label{fig:BaryondensityT}
\end{figure}

For temperatures higher than the one of the critical endpoint $T_*=20.7\;\text{MeV}$ one finds that the first order phase transition gets replaced by a crossover which gets rapidly rather smooth. To illustrate this we plot in Fig.\ \ref{fig:sigmamin} the chiral order parameter $\sigma_0$ as a function of the baryon chemical potential $\mu$ for various values of the temperature in the range $T=0\dots 80\;\text{MeV}$. Similarly, Fig.\ \ref{fig:BaryondensityT} shows the baryon density $n_B$ as a function of the chemical potential for the same temperatures. Together with Fig. \ref{fig:2} and Fig. \ref{fig:3} this demonstrates clearly that the observed chemical freeze-out points in the $\mu-T$-diagram are far away from any phase transition or rapid crossover. This main result of the present letter contrasts with the properties of chemical freeze-out at low baryon density.

Still, the baryon density (or particle density if mesons are included) changes rapidly as a function of temperature in the range of interest, cf. fig. \ref{fig:2}. The rates of processes with a change of particle numbers, as annihilation or production of strange particles, depends strongly on the number density of hadrons \cite{BMSW}. These processes stop effectively once the hadron density drops below a critical value. Since a rather small change of temperature corresponds to a substantial change in hadron density the chemical freeze-out occurs for a small temperature interval. This explains why common values of $\mu$ and $T$ can describe rather well the abundancies of all hadron species. Even if the freeze-out number density for different hadrons differs to some extent, this will not have a large effect on the value of the freeze-out temperature.

We plot in Fig. \ref{fig:1} the line of constant baryon density, $n=0.15\;n_\text{nuclear}$. The observed freeze-out temperatures are well described by this line if the baryon density is high enough. Let us remark, however, that the true baryon density at chemical freeze-out might be somewhat larger due to the contribution of Delta baryons which we have neglected in our model. %For a discussion of baryon number densities within the statistical model and a polynomial parameterization of the chemical freeze-out line see refs.\ \cite{RandrupCleymans}. 

\begin{figure}
\centering
\includegraphics[width=0.5\textwidth]{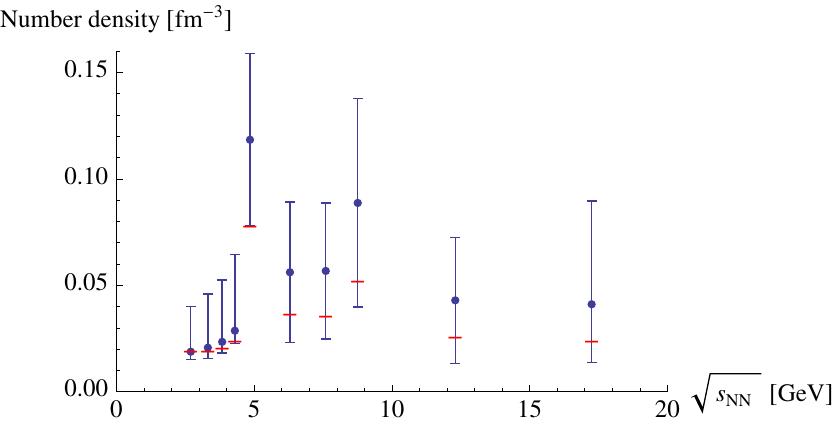}
\caption{Contribution of protons, neutrons and delta baryons to the baryon number density as a function of the center of mass energy. These densities have been calculated in the statistical model based on occupation numbers for free particles and resonances and using the chemical potentials end temperatures from the fits to particle number ratios performed in ref.\ \cite{ABMS2009}. The error bars correspond to the uncertainty in the estimation of the freeze-out temperature. We also indicate the baryon number density due to protons and neutrons only (red bars).}
\label{fig:BaryondensityExperiments}
\end{figure}

To test the hypothesis of a chemical freeze-out at constant baryon number density further we plot in Fig.\ \ref{fig:BaryondensityExperiments} the baryon number densities as calculated for the statistical model (occupation numbers of free particles and resonances with vacuum masses and decay widths) corresponding to the values of chemical potentials and temperatures fitted to particle number ratios in ref.\ \cite{ABMS2009}. In this representation it becomes apparent that the experimental results for $\sqrt{s_{NN}}\lesssim 4.5\;\text{Gev}$ are in good agreement with the above presumptions and that the contribution from Delta baryons is not too large in this regime.

We note that in previous studies \cite{RandrupCleymans, Cleymansetal} the baryon number density at freeze-out was partly overestimated. These studies were based on parameterizations of the freeze-out curve in the $T$-$\mu$-plane, which seem to work satisfactory there but give a rather poor account of the experimental results with respect to baryon number densities. The problem is that small changes in the $T$-$\mu$-plane can lead to strong changes of the baryon number density in the relevant regime.

We observe that any line with $n=c\, n_\text{nuclear}$, with $c<1$, will end for low $T$ in the line of the gas-liquid nuclear phase transition. This simply follows from the jump of $n$ from $n=0$ to $n=n_\text{nuclear}$ for $T=0$. This explains why the observed freeze-out points tend to approach the first order nuclear phase transition. As we have seen, however, this should not be interpreted as a continuation of this first order line by a strong crossover. 

We note as an aside that a freeze-out at constant baryon number density can explain also a puzzling feature of the chemical freeze-out volume per unit rapidity $dV/dy$. While this quantity grows with collision energy for the high energy experiments, the relation is reverse for $\sqrt{s_{NN}}\lesssim 4.5\;\text{Gev}$ \cite{ABMS2006}. However, assuming constant $n_\text{Baryons}$ at freeze-out, the behavior at AGS energies reflects just the experimental finding for the energy dependence of the proton yield $dN/dy$ at $y=0$ \cite{AGS}.

In principle, a deviation of the chiral condensate $\sigma_0$ from its vacuum value $f_\pi$ and a non-vanishing expectation value $\omega_0$ (implying $\mu_\text{eff} = \mu+g \omega_0 \neq \mu$) should lead to modified particle yields with respect to a thermal model that assumes vacuum masses and $\mu=\mu_\text{eff}$. The chemical potential extracted from the thermal fits corresponds in first approximation actually to $\mu+g \omega_0-h(\sigma_0-f_\pi)$. However, we find the deviation of this quantity from $\mu$ to be small in the relevant regime, typically $\sim 10\;\text{MeV}$.

\begin{figure}
\centering
\includegraphics[width=0.5\textwidth]{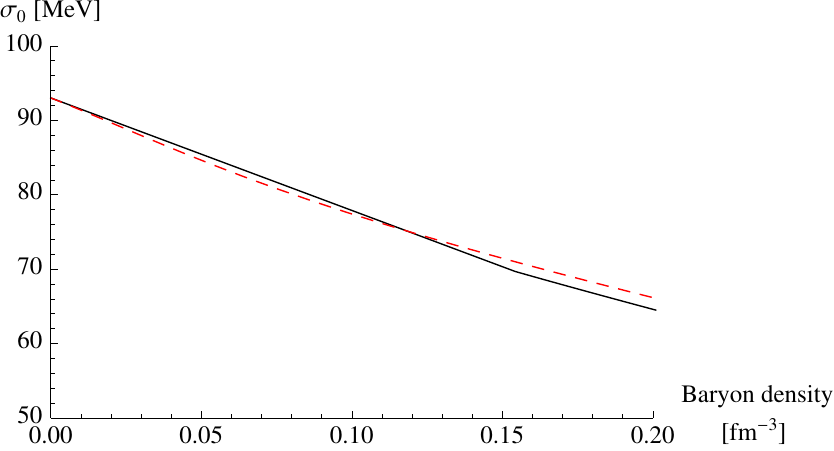}
\caption{Chiral oder parameter $\sigma_0$ as a function of the baryon density for vanishing temperature $T=0$ (solid line) and for $T=50\;\text{MeV}$ (dashed line).}
\label{fig:SigmaBaryondensity}
\end{figure}

The information contained in Figs.\ \ref{fig:sigmamin} and \ref{fig:BaryondensityT} can be combined to yield the chiral condensate as a function of the baryon density. In Fig.\ \ref{fig:SigmaBaryondensity} we show this dependence for the temperatures $T=0$ and $T=50\;\text{MeV}$. At vanishing temperature we find a rather good quantitative agreement with calculations based on in-medium chiral perturbation theory \cite{CPT2}. However, whereas our results show only a rather weak temperature dependence, the calculation performed in ref.\ \cite{CPT3} shows stronger deviations. It is found there that the chiral condensate decreases faster with the baryon density at $T=50\;\text{MeV}$.

We may also explore the phase diagram for temperatures higher than the freeze-out temperature. The temperature dependence of the number density and energy density for various values of $\mu$ is shown in figs. \ref{fig:Numberdensitymu} and \ref{fig:Energydensitymu}.

\section*{Meson fluctuations}
For our quantitative discussion we have not included the fluctuations of the $\pi,\sigma$, and $\omega$ mesons. We also did not consider their contribution to the particle density for hadrons. For an estimate of the validity of our approximation we may compute the contribution of the meson fluctuations to the effective potential, using a Gaussian approximation similar to the baryon fluctuations. For that purpose we assume that the momentum dependent parts of their inverse propagators are not modified by the effects of non-zero chemical potential and temperature, and we use a Gaussian or quasi-particle approximation. The effect of the bosonic fluctuations can then be parametrized in terms of the pressure of a free gas of relativistic bosons, $P_\text{BG}(T,\mu,m)$, completely analogous to Eq.\ \eqref{eq:PFG}. The masses of the pions and $\sigma$-mesons depend on $\sigma$ according to
\begin{eqnarray}\label{16}
m^2_\pi(\sigma)=\frac{1}{\sigma} \left(\frac{\partial U}{\partial\sigma}+m^2_\pi f_\pi\right)~,~
m^2_\sigma(\sigma)=\frac{\partial^2 U}{\partial\sigma^2},
\end{eqnarray}
while $m^2_\omega$ is independent of $\sigma$ in our approximation. Thus the $\omega$-fluctuations contribute to $U$ only a temperature dependent constant and we will neglect them. The Gaussian approximation becomes invalid in regions where $m^2_\pi(\sigma)$ or $m^2_\sigma(\sigma)$ become negative. Close to the value where $m^2_\pi(\sigma)$ vanishes it also matters if one evaluates $\partial U/\partial\sigma$ for the potential \eqref{eq:effpotentialcomb}, as we do it here, or uses a self-consistent formulation where $U$ includes the meson fluctuations. We stay away from such problematic regions. 

\begin{figure}
\centering
\includegraphics[width=0.5\textwidth]{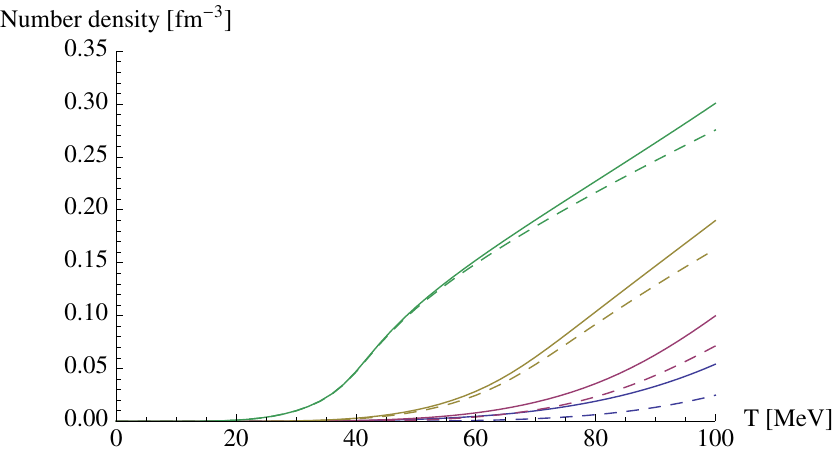}
\caption{Number density of baryons and pions (solid lines) as well as baryons only (dashed lines) as a function of temperature for the chemical potentials $\mu=550\;\text{MeV}$ (lowermost curves), $\mu=650\;\text{MeV}$, $\mu=750\;\text{MeV}$ and $\mu=850\;\text{MeV}$ (uppermost curves).}
\label{fig:Numberdensitymu}
\end{figure}

\begin{figure}
\centering
\includegraphics[width=0.5\textwidth]{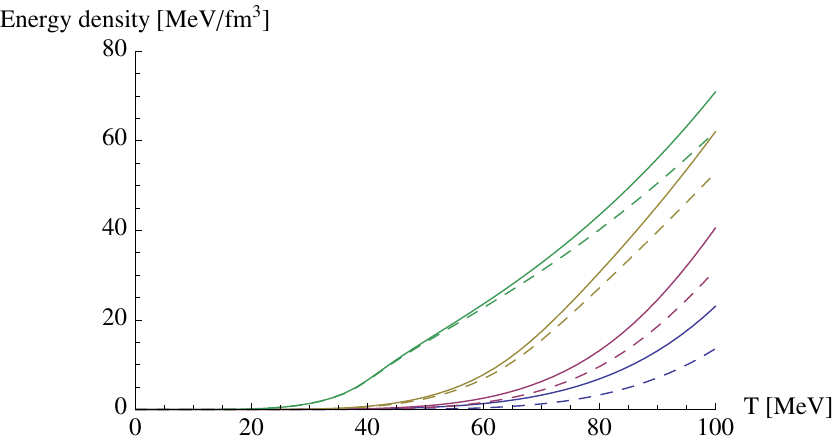}
\caption{Energy density of baryons and pions (solid lines) as well as baryons only (dashed lines) as a function of temperature for the chemical potentials $\mu=550\;\text{MeV}$ (lowermost curves), $\mu=650\;\text{MeV}$, $\mu=750\;\text{MeV}$ and $\mu=850\;\text{MeV}$ (uppermost curves).}
\label{fig:Energydensitymu}
\end{figure}

The pion- and $\sigma$-contributions to the effective potential $\Delta U_m$ read 
\begin{equation}
\Delta U_m=
- 3 P_\text{BG}\left(T, 0, m_\pi(\sigma) \right)
-  P_\text{BG}\left(T, 0, m_\sigma(\sigma) \right).
\label{eq:fullU}
\end{equation}
Here $P_{BG}$ obtains from Eq. \eqref{eq:PFG} by setting $\mu=0$, dropping the second term and replacing the fermionic by the bosonic mean occupation number (changing $+1$ to $-1$ in the denominator). We have verified that the meson fluctuations play only a minor role in the region of interest, as shown in Fig. \ref{fig:1} and \ref{fig:2}. The relative importance of the pion fluctuations can be judged from figs. \ref{fig:Numberdensitymu} and \ref{fig:Energydensitymu}.

We conclude that the central result of this note seems to be rather robust. The chemical freeze-out at high baryon density is not related to any phase transition or rapid crossover. It rather follows a simple line of constant freeze-out density. We believe that the linear nucleon-meson model can give a reliable description for the whole low temperature region of the QCD-phase diagram up to densities of at least $1.5$ times nuclear density. This requires the inclusion of the meson fluctuations, for example by a genuine functional renormalization group study. 

\section*{Acknowledgment}
S.~F.~acknowledges useful discussions with A.~Andronic and financial support by DFG under contract FL 736/1\hbox{-}1.

\end{document}